\documentclass[%
 aps,
 prb,%
 amsmath,amssymb,
preprint,showpacs,showkeys,%
]{revtex4-1}

\usepackage{graphicx}
\usepackage{dcolumn}
\usepackage{bm}

\begin{document}


\title{Analytical model to predict the effect of a finite impedance surface on the propagation properties of a 2D Sonic Crystal}

\author{V. Romero-Garc\'ia}
 \email{virogar1@mat.upv.es}
 \affiliation{Instituto de Ciencia de Materiales de Madrid, Consejo Superior de Investigaciones Cient\'ificas}
 \altaffiliation{Centro de Tecnolog\'ias F\'isicas: Ac\'ustica, Materiales y Astrof\'isica, Universidad Polit\'ecnica de Valencia.}
\author{J.V. S\'anchez-P\'erez}
 \affiliation{Centro de Tecnolog\'ias F\'isicas: Ac\'ustica, Materiales y Astrof\'isica, Universidad Polit\'ecnica de Valencia.}

\author{L.M. Garcia-Raffi}
 \affiliation{Instituto Universitario de Matem\'atica Pura y Aplicada, Universidad Polit\'ecnica de Valencia.}

\date{\today}

\begin{abstract}

The use of Sonic Crystals as environmental noise barriers has certain
advantages from the acoustical and the constructive point of view
with regard to conventional ones. One aspect do not studied yet is the
acoustic interaction between the Sonic Crystals and the ground due
to, up to now,  this latter is not included in the analytical models
used to characterize these Sonic Crystals. We present here an
analytical model, based on multiple scattering theory, to study this
interaction considering the ground as a finite impedance surface. Using this model we have obtained interesting
conclusions that allow to design more effectively noise screens
based on Sonic Crystals. The obtained results have been compared
with experimental and numerical, finding a good agreement between
them.
\end{abstract}


\maketitle

\section{Introduction}
Periodic arrangement of acoustic scatterers embedded in a medium
with different physical properties give rise to ranges of
frequencies, known as bandgaps, where the transmission of acoustic
waves is forbidden. If one of them, scatterers or host, is a fluid then
these systems are called Sonic Crystals (SC). In last years, an increasing interest has appeared in the potential exploitation of SC as environmental noise barriers \cite{Sanchez-PerezA, Sanchez10, SUOU}.

Some examples of the advantages of using SC instead of conventional screens are the reduction in the size of the foundation or the possibility of designing specific screens for predetermined conditions. However, the acoustical properties of SC depend on several factors showing some particularities in their attenuation properties \cite{Sanchez-PerezB}. For example, the size and position of the bandgaps depend on several factors as the direction of incidence of the wave on the SC and the type of arrangement of the scatterers. As a consequence, the development of the screens based on SC is not a trivial process.

In order to avoid these handicaps several works have been intensively
developed in last years. The use of both materials with acoustical
properties added or more efficient distribution of scatterers are two examples. The use
of resonators \cite {Liu00a} or absorbent materials
\cite{Umnova} in the first case, and the use of Quasi Ordered
\cite{Romero06} or Quasi Fractal \cite{Castineira10} structures
in the second case have been studied.

One of the factors to have into account in the use of SC as noise
barriers is the existence of the ground. Up to now, one of the analytical approaches to predict the transmission properties of SC is based on the well known Multiple Scattering Theory (MST) \cite{Zaviska, Linton, Martin, Chen01}, which is a self-consistent method for calculating acoustic pressure including all orders of scattering based on the superpositions of the solution for a simple scatterer. MST predicts the acoustical performance of SC in the absence of a ground plane. However, one of the factors to have into account in the use of SC as noise barriers is precisely the existence of the ground. Thus MST is an
unrealistic approximation for environmental noise barriers based on SC.

In this work, the effect of both acoustically-rigid and finite impedance ground planes on the properties of a SC made of rigid scatterers is analytically and experimentally analyzed. The methodology developed in this work is based on the MST and on the method of images \cite{Allen79, Boulanger}. The finite impedance is characterized in the model as a two parameter impedance \cite{Taherzadeh}. Although the
most interesting situation is likely to involve periodic vertical
cylinder arrays, this would require solution of a 3D problem. Here
are considered the more tractable 2D problem involving a periodic
array of cylinders with their axes parallel to the ground.

\subsection{Defining the problem}
\begin{figure}
\center
    \includegraphics{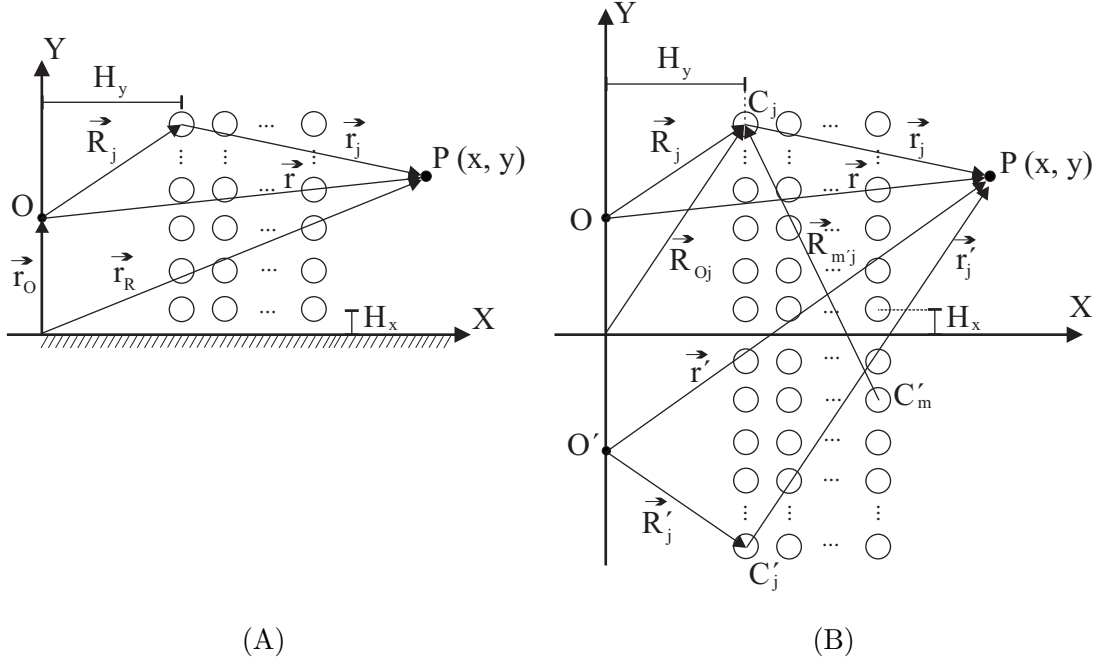}\\
    (A)\hspace{7cm}(B)
    \caption{(A) Square lattice array above a perfectly reflecting plane. (B) Schematic of the real and the image sources and the real and the image scatterers.}
     \label{fig:geom}
\end{figure}

The problem studied in this work is related to the scattering of sound waves by an array of scatterers suspended with their axes parallel to a rigid or finite impedance ground plane. As it is shown in Figure \ref{fig:geom}A, the scattering is produced in the positive half-space.

Consider a line source placed at point $O$ and an array of $M$ circular scatterers placed in the positive half-space which is air, characterized by the sound velocity, $c=344$ m/s, and density $\rho=1.23$ $kg/m^3$. The position of each scatterer $C_m,\,m=1,\ldots,M$ is given by the vector $\vec{R}_m$. Figure \ref{fig:geom}A shows an scheme of the problem. The scatterers are considered to be arranged in a square lattice which is defined by the lattice constant $a$. The nearest base of the array of scatterers is placed at a distance $H_x$ from the ground, while the nearest vertical base of the array from the source is placed at a distance $H_y$ (see Figure \ref{fig:geom}A)

Here we have studied the effect of the ground over the scattering of the array of scatterers by means of the multiple scattering theory modified using the method of images in order to construct the reflected field. The geometry used to perform the method of images in our approach is shown in Figure \ref{fig:geom}B. In this approach one should consider the image of the source as well as of the scatterers. Note that all the waves reflected on the ground can be described as waves coming from the image source or from the image scatterers, then the images are also interacting with the real space. The image source is placed at point $O'$, and the image of scatterers $C'_m$ are placed on the negative half-space. All the vectors measured from the image source is characterized by a prime ($'$). 

\subsection{Ground effect}
\label{sec:ground}
In some cases the surface of the ground can be considered perfectly rigid or totally reflective, however there are some cases, as for example open water, ice, concrete or soil surfaces
with or without vegetation, in which there is absorption of energy from incident acoustic waves. Accurate
prediction of ground effects on the scattering of an array of scatterers requires knowledge of the absorptive and reflective properties (the
acoustic impedance) of the surface. Motivated by previous works \cite{Allen79} the effect of the ground on the scattering problem will be characterized in this work by the reflection coefficient $R(\vec{r}_O,\vec{r}_R;\nu)$ of the ground ($\nu$ is the frequency of the incident wave). The case of an acoustically-hard (rigid) ground implies $R(\vec{r}_O,\vec{r}_R;\nu)=1$. In general, $R(\vec{r}_O,\vec{r}_R;\nu)$  will be a function that mainly depends on the impedance contrast between the two half-spaces separated by the ground surface and on the positions of both the source ($\vec{r}_{O}$) and the receiver ($\vec{r}_{C}$) by means of the angle of incidence on the ground. 

The ground surface itself also provides a significant path for transmission of acoustic energy,
particularly at low grazing angles and low frequencies. Incident acoustic energy is transformed
into vibrational energy and is transmitted along the surface layer. This vibration disturbance can
propagate for long distances, before dissipating or re-radiating as sound. At long distances, the
transmission of low frequency sound can be dominated by this surface wave mechanism. In this work we are interested in the interaction between the SC and the ground effect, to do that we study regions near the source (the positions of both the array of scatterers and the receiver are near the source) and in the regime of dispersion frequencies of the array (high frequencies), this transmission mechanism can be neglected in this work.

When airborne sound impinge on the ground, a portion of the wave is transmitted and refracted at right
angles into the surface. For our purposes, we have focused our attention on the reflected portion of the incident wave. The reflected wave leaves the surface at the angle of
incidence, with its amplitude and phase modified by the impedance of the surface. The reflected
wave propagates to the receiver, in addition to the direct wave from the source. Depending on their
relative phases and amplitudes, they may constructively add or destructively interfere \cite{Piercy77, Attenborough92, Attenborough95}. The effect of the ground on the propagating wave from the source in the receiver site is an increasing of the attenuation usually called excess attenuation. This can be explained in terms of the existence of two sources: the real one and the image source that model the reflected wave. In this case, the governing equation for the pressure amplitude $p$ at the receiver, assuming a uniform medium (no refraction) and line source, in the positive half-space is
\begin{align}
\label{eq_amplitude}
p=&H_0(kr)+R(\vec{r}_O,\vec{r}_R;\nu)H_0(kr')=\nonumber\\
&H_0(kr)+R_p(\vec{r}_O,\vec{r}_R,\nu)H_0(kr')+(1-R_p(\vec{r}_O,\vec{r}_R,\nu))FH_0(kr')
\end{align}
where $R_p$ is the plane wave coefficient, $H_0$ is the Hankel function of 0-th order and first kind. The parameter $F$ is the boundary-loss factor which is a complicated mathematical function of a variable ${\rm w}$ called the numerical distance. These functions are \cite{Piercy77, Taherzadeh}
\begin{eqnarray}
\label{eq:R}
R_p(\vec{r}_O,\vec{r}_R,\nu)=\frac{\cos\theta-\frac{Z_{air}}{Z_{ground}}}{\cos\theta+\frac{Z_{air}}{Z_{ground}}}\\
\label{eq:F}
F=1+\imath\sqrt{\pi }{\rm w}e^{-{\rm w}^2}{\rm erfc}(-\imath {\rm w}),
\end{eqnarray}
where
\begin{equation}
\label{eq:w}
{\rm w}=\sqrt{\frac{1}{2}\imath k r_2}\left(\cos\theta+\frac{Z_{air}}{Z_{ground}}\right),
\end{equation}
$Z_{air}$ and $Z_{ground}$ are the air and ground impedance respectively, $r_2$ is the distance between the reflection point and the receiver and $\theta$ is the reflection angle measured from the normal of the surface. ${\rm erfc}$ is the complex complementary error function. Usually, the fraction $\beta=Z_{air}/Z_{ground}$ is called the admittance of the homogeneous impedance plane. The reflected angle can be obtained as
\begin{equation}
\label{eq_theta}
\theta=arctan\left(\frac{x-x_0}{y+y_0}\right)
\end{equation}
where $\vec{r}_0=(x_0,y_0)$ is the position of the source and $\vec{r}_R=(x,y)$ is the position of the receiver point with respect to the origin of coordinates (see Figure \ref{fig:geom}).

Note that for acoustically-rigid ground, as $Z_{ground}>>Z_{air}$, then $R(\vec{r}_O,\vec{r}_R;\nu)=1$. For the case of the finite impedance ground, it is necessary to know the expression of the impedance in order to know the reflection properties of the surface. In this work the finite impedance (open cell foam layer) surface is represented by a two parameter impedance model \cite{Taherzadeh} with flow resistivity $\sigma_e = 4\,{\rm kPa\,s/m^2}$ and porosity at the surface $\alpha_e = 105\,{\rm m^{-1}}$, being the impedance of the ground
\begin{equation}
\label{impedance}
Z_{ground}=\rho c_0\left(0.434\sqrt{\frac{\sigma_e}{\nu}}(1+\imath)+9.75\imath\frac{\alpha_e}{\nu}\right).
\end{equation}

\section{Image multiple scattering theory (IMST)}
\label{IMST}
The solution of the appropriate scattering problem  satisfies the Helmholtz equation in the 
half-space that is written in polar coordinates $(r,\theta)$ as
\begin{equation}
\label{form_Helmholtz}
	\Delta p(\vec{r}) + k^2 p(\vec{r})=0,
\end{equation}
where $\displaystyle{\Delta = \frac{1}{r}\frac{\partial}{\partial r}\left(r\frac{\partial}{\partial r}\right)+\frac{1}{r^2}\frac{\partial^2}{\partial \theta^2}}$, $\vec{r}=r(\cos\theta,\sin\theta)$ is the radius vector, $p$ is acoustic displacement potential, $k=\omega/c$ and $\omega$ 
is angular frequency. Equation \ref{form_Helmholtz} is solved in conjunction with radiation conditions
\begin{equation}
\label{form_radiation}
	\frac{\partial p}{\partial r} - i k p = o\left(r^{-1/2}\right),\;\text{as}\; r \rightarrow \infty.
\end{equation}

Given $M$ disjoint cylindrical scatterers located at the positions $\vec{R_m}=\vec{R}_1,...,\vec{R}$ all placed above a reflectance surface on the symmetry axis (see Figure \ref{fig:geom}) and a sound source located at point $O$, one can consider the incident field over the $n$-th scatterer in the presence of both the other $ M-1$ scatterers and the ground by taking into account the multiple scattering technique \cite{Zaviska, Linton} as well as the method of images \cite{Allen79, Boulanger}
\begin{equation}
\label{form_tfield}
	p^n(\vec{r})= p_0(\vec{r}) + \sum_{j=1,j\neq n}^M \left(p_s^j(\vec{r})+R(\vec{R}_{Oj},\vec{r}_R;\nu) p_s^{j'}(\vec{r})\right),
\end{equation}
where $p_0$ is the pressure produced by both the real and the image sources and $p_s^j$ and $p_s^{j'}$ are the scattered pressure by the $j$-th cylinder and its image $j'$-th cylinder respectively. Equation \ref{form_tfield} defines the interaction between the scattering of the array and the ground, therefore the analytical method shown in this work is called Image multiple scattering theory (IMST). The pressure of the sources can be expressed as
\begin{equation}
\label{eq_p_0}
	p_0(\vec{r})= H_0(kr)+R(\vec{r}_O,\vec{r}_R;\nu)H_0(kr'),
\end{equation}
where $\vec{r}=r(\cos\theta, \sin\theta)$ is the vector connecting the real line source and the receiver point and $\vec{r'}=r'(\cos\theta', \sin\theta')$ connects the position of the receiver and the image point source. On the other hand, the scattered pressures by the cylinders $m$ and $m'$ can be represented as
\begin{eqnarray}
\label{eq_p_sc}
	p^m_s(\vec{r})= \sum_{l=-\infty}^{\infty}A^m_lH^{(1)}_l(kr_m)e^{\imath  l\theta_m},\\
\label{eq_p_sc_prima}
	p^{m'}_s(\vec{r})= \sum_{l=-\infty}^{\infty}A^{m'}_lR(\vec{R}_{Om},\vec{r}_R;\nu)H^{(1)}_l(kr_{m'})e^{\imath  l\theta_{m'}},
\end{eqnarray}
where $\vec{r}_m = r_m (\cos\theta_m, \sin\theta_m)$ is vector connecting the centre of scatterer $C_m$ and the receiver and $\vec{r}_{m'} = r_{m'} (\cos\theta_{m'}, \sin\theta_{m'})$ connects the receiver with the scatterer image $C_{m'}$. Note that the reflected wave by the cylinder image, is produced by the presence of the ground, thus the reflected pressure by the cylinder should be modulated by the reflection coefficient of the ground as in the case of the incident wave on the ground.
 
In order to introduce Equations \ref{eq_p_0}, \ref{eq_p_sc} and \ref{eq_p_sc_prima} in Equation \ref{form_tfield} all the terms must be expressed in the same origin of coordinates. To do so, the Graf's addition theorems for the Bessel and Hankel functions are necessary \cite{Martin, Abramowitz}. 
Thus, the pressures $p_0$ and $p_s$ in the reference system centered at $n-$th scatterer are,
\begin{eqnarray}
\label{eq_p_0_and_p_s_1}
p_0(\vec{r})=\sum_{l=-\infty}^{\infty}(H^{(1)}_{-l}(kR_n)e^{-\imath l\theta_{R_n}}+R(\vec{r}_O,\vec{r}_R;\nu)H^{(1)}_{-l}(kR'_n)e^{-\imath l\theta_{R'_n}})J_l(kr_n)e^{\imath l\theta_n}\\
\label{eq_p_0_and_p_s_2}
p^j_s(\vec{r})=\sum_{l=-\infty}^{\infty} (G^{jn}_m+R(\vec{R}_{Oj},\vec{r}_R;\nu)G^{j'n}_m)J_l(kr_n)e^{\imath l\theta_n}\\
G^{jn}_m=\sum_{s=-\infty}^{\infty}A^j_s H^{(1)}_{m-s}(kR_{jn})e^{\imath (m-s)\theta_{jn}}=\sum_{s=-\infty}^{\infty}A^j_sG^{jn}_{ms}\\
G^{j'n}_m=\sum_{s=-\infty}^{\infty}A^{j'}_s H^{(1)}_{m-s}(kR_{j'n})e^{\imath (m-s)\theta_{j'n}}=\sum_{s=-\infty}^{\infty}A^{j'}_sG^{j'n}_{ms}\\
\end{eqnarray}
where vector $\vec{R}_{n}=R_{n}(\cos\theta_{R_n}, \sin\theta_{R_n})$ ($\vec{R}'_{n}=R_{n}(\cos\theta_{R'_n}, \sin\theta_{R'_n})$) defines the position of scatterer $C_n$ with respect to real (image) line source and vector $\vec{R}_{jn}=R_{jn}(\cos\theta_{jn}, \sin\theta_{jn})$ ($\vec{R}_{j'n}=R_{j'n}(\cos\theta_{j'n}, \sin\theta_{j'n})$) defines the position of scatterer $C_j$ $(C_{j'})$ with respect to scatterer $C_n$.

Finally, because of the geometry of the problem, we can express the total incident wave over the $n$-th scatterer as
\begin{equation}
\label{eq_p}
	p^n(\vec{r})= \sum_{s=-\infty}^{\infty} B^n_s J_s(k r_n)e^{\imath s\theta_n}.
\end{equation}
Introducing Equation \ref{eq_p_0_and_p_s_1}, \ref{eq_p_0_and_p_s_2} and \ref{eq_p} in Equation \ref{form_tfield}, one can obtain the following system of equations,
\begin{equation}
\label{eq:system}
B^n_s=S^n_s+\sum_{j=1}^{M}\left((1-\delta_{jn})G^{jn}_s+R(\vec{R}_{Oj},\vec{r}_R;\nu)G^{j'n}_m\right),
\end{equation}
where
\begin{equation}
S^n_s=H^{(1)}_{-l}(kR_n)e^{-\imath l\theta_{R_n}}+R(\vec{r}_O,\vec{r}_R;\nu)H^{(1)}_{-l}(kR'_n)e^{-\imath l\theta_{R'_n}}.
\end{equation}
At this stage $B^n_s$, $A^{j}_s$ and $A^{j'}_s$ are unknown coefficients but they can be related using the boundary conditions on the scatterers and the symmetry of the problem. The boundary conditions at a rigid cylinder surface relates $B^j_s$ with $A^j_s$ and the symmetric condition
relates $A^j_s$ with $A^{j'}_s$. In our approach we will consider the general
boundary condition, i.e., the continuity of both the pressure and
the normal velocity across the interface between the scatterers and
the surrounding medium. After that, considering the big contrast
between both the densities and sound velocities, it will be possible to reproduce the results of rigid
scatterers (Neumann boundary conditions).

The boundary conditions in the $n$-th scatterer can be expressed as
\begin{eqnarray}
\label{eq:continuous_11}
p_{ext}|_{\partial \Omega_n}=p_{int}|_{\partial \Omega_n},\\
\label{eq:continuous_21} \frac{1}{\rho}\frac{\partial
p_{ext}}{\partial n}\rvert_{\partial \Omega_n
}=\frac{1}{\rho_n}\frac{\partial p_{int}}{\partial
n}\rvert_{\partial \Omega_n},
\end{eqnarray}
where $\partial \Omega_n$ is the boundary of the $n$-th scatterer,
$\rho$ is the density of the surrounding medium and $\rho_n$ is
the density of the $n$-th scatterer.

In order to apply the previous boundary conditions, we consider
that the pressure field inside the $n$-th cylinder can be
represented by
\begin{eqnarray}
P_{int}^n=\sum_{j=-\infty}^{\infty} D_{j}^n J_j(k_{1n}r_n)e^{\imath 
j \theta_n},
\end{eqnarray}
where $k_{1n}$ is the wave number inside the $n$-th cylinder.

Using the boundary conditions and the expressions of both the exterior and interior fields in the $n-$th scatterer, we can obtain the following relation,
\begin{eqnarray}
\label{eq:coef}
B_{j}^n=\Gamma_{j}^nA_{j}^n,
\end{eqnarray}
where
\begin{eqnarray}
\label{eq:Gamma}
\Gamma_{j}^n=\frac{H_j(ka_n)J_n'(ka_n/h_n)-g_nh_nH_j'(ka_n)J_j(ka_n/h)}{g_nh_nJ_j'(ka_n)J_j(ka_n/h_n)-J_j(ka_n)J_j'(ka_n/h_n)}.
\end{eqnarray}
Here $a_n$ is the radius of the $n$-th cylinder (in this work the radius of the scatterers take the same value for all the cylinders, $a_n=a$),
$g_n=\rho_1^n/\rho$ is the density ratio, and
$h_n=k/k_1^n=c_1^n/c$ is the sound speed ratio for the $i$-th
cylinder. Note that if the scatterers are acoustically hard, i.e., $\rho_1>>\rho$ and $c_1>>c$, then the coefficients $\Gamma_j^n$ coincides with those obtained with the Neumann boundary conditions,
\begin{eqnarray}
\label{eq:Gamma_rigid}
\Gamma_{j}^n=-\frac{\partial_r H_j(ka_n)}{\partial_r J_j(ka_n)},
\end{eqnarray}
where $\partial_r$ is the derivative with respect to polar coordinate $r$.

The image symmetry can be used to relate $A^j_s$ with $A^{j'}_s$. One have to take into account that $r_{j'}=r_{j}$ and that $\theta_{j'}=-\theta_{j}$, then
\begin{align}
p^{j'}_s(\vec{r})= &R(\vec{r}_O,\vec{r}_R;\nu)\sum_{l=-\infty}^{\infty}A^{j'}_lH^{(1)}_l(kr_{j'})e^{\imath  l\theta_{j'}}\nonumber\\
&=R(\vec{R}_{Oj},\vec{r}_R;\nu)\sum_{l=-\infty}^{\infty}A^{j}_lH^{(1)}_l(kr_{j})e^{-\imath  l\theta_{j'}}\nonumber\\
&=R(\vec{R}_{Oj},\vec{r}_R;\nu)\sum_{l=-\infty}^{\infty}A^{j}_{-l}H^{(1)}_{-l}(kr_{j})e^{\imath  l\theta_{j'}}\nonumber\\
&=R(\vec{R}_{Oj},\vec{r}_R;\nu)\sum_{l=-\infty}^{\infty}A^{j}_{-l}(-1)^{l}H^{(1)}_{l}(kr_{j'})e^{\imath  l\theta_{j'}},
\end{align}
and
\begin{equation}
A^{j'}_l=(-1)^lA^j_{-l}.
\end{equation}

Introducing the Equation \ref{eq:Gamma} or \ref{eq:Gamma_rigid} in \ref{eq:coef} and in \ref{eq:system}, the following infinite system of equations is obtained,
\begin{equation}
\label{eq:system_1}
\Gamma_{s}^nA_{s}^n=S^n_s+\sum_{j=1}^{M}\sum_{l=-\infty}^{\infty}\left((1-\delta_{jn})G^{jn}_{sl}+(-1)^lR(\vec{R}_{Oj},\vec{r}_R;\nu)G^{j'n}_{s -l}\right)A^n_l.
\end{equation}
The coefficient $A_{s}^n$ can be obtained by truncating properly the previous system, and the total acoustic field obtained using the IMST is
\begin{align}
P(\vec{r})&=H_0(kr)+R(\vec{r}_O,\vec{r}_R;\nu)H_0(kr')\nonumber\\
\label{eq:total_pressure}
&+\sum_{m=1}^M\sum_{l=-\infty}^{\infty}A^m_l\left(H^{(1)}_l(kr_{m})e^{\imath  l\theta_{m}}+R(\vec{R}_{Oj},\vec{r}_R;\nu)H^{(1)}_l(kr_{m'})e^{-\imath  l\theta_{m'}}\right).
\end{align}
Note that the methodology is self-consistent and the effect of the finite impedance of the ground only depends on the model to calculate this impedance, i.e., on the calculation of the reflection coefficient. The two parameter impedance model used in this work \cite{Taherzadeh} is constrained for low grazing angles, however, due to both the simplicity of the model and the good agreement with the experimental data (see Section \ref{comparisons}) we have decide use it. In this sense IMST is completely general because any impedance model can be used. 

When the real source is placed on the origin of coordinates ($O=(0,0)$), then the image source coincides with the real one and depending on the properties of the ground several interesting possibilities can be analyzed. With the source on an aocustically-hard ground, the predicted IL spectrum of an array, for instance, $5\times3$ array in the presence of the rigid ground ($R(\vec{r}_O,\vec{r}_R;\nu)=1$) should be the same as that predicted for an array of double the size ($10\times3$) in the free field, whereas in presence of an completely absorbent ground ($R(\vec{r}_O,\vec{r}_R;\nu)=0$) the IL should be the same as a  the initial array, $5\times3$, in free field. 

\begin{figure}[h]
		\center
		\includegraphics[scale=.43]{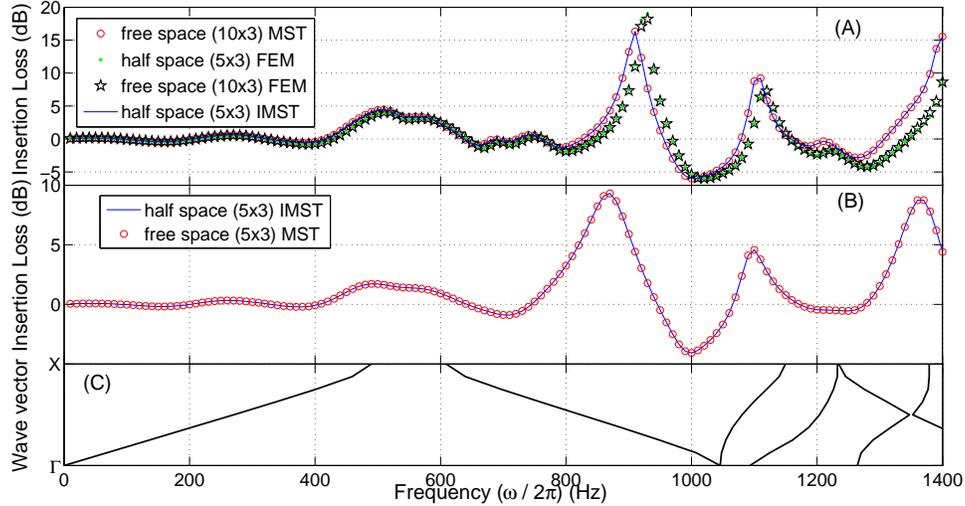}
    \caption{(Color online) Predicted IL spectra of a square array of rigid cylinders with diameter 0.1 m, lattice constant 0.3 m above ground plane at y = 0 and 'lowest' cylinder centers at $H_x = 0.15$ m. The source and receiver coordinates are (0,0) m and (2.5,0) m respectively. The nearest part of the array is at $H_y=1$ m from the source. (A) Acoustically-hard ground $R(\vec{r}_O,\vec{r}_R, \nu)=1$: Blue continuous line (Green dots) shows the IL produced by a $5\times3$ square array of rigid cylinders calculated using IMST (finite element method). Red open circles (Black pentagrams) show the IL produced by a doubled (i.e. a $10\times3$) lattice in the free field calculated using MST (finite element method). (B) Competelly absorbent ground $R(\vec{r}_O,\vec{r}_R, \nu)=0$: Blue continuous line (Open red circles) shows the IL produced by a $5\times3$ square array of rigid cylinders calculated using IMST (MST). 
(C) Band structure calculated using plane wave expansion (PWE) for this case.}
    \label{fig:formulations_comparison1}
\end{figure}

We analytically and numerically compare the predicted insertion loss (IL) spectrum due to a $5\times3$ array of rigid scatterers considering two particular grounds: an acoustically-rigid ground characterized by $R(\vec{r}_O,\vec{r}_R;\nu)=1$ and a completely absorbent ground $R(\vec{r}_O,\vec{r}_R;\nu)=0$. Although, the last case is physically impossible it could be used to test the model with the equivalent situation without ground. 
For these simulations we have chosen a the lattice constant $a = 0.3$ m and a diameter of scatterers of 0.1 m. Note that throughout this paper the IL is calculated as
\begin{equation}
\label{form_IL}
	IL=20\log_{10}\left|\frac{p_0}{P}\right|.
\end{equation}
where $P$ is calculated using Equation \ref{eq:total_pressure}.
The distance to the ground of the centers of the lowest cylinders in the array is half of the lattice constant, $H_x=0.15$ m, so that they are separated from the centers of the cylinders of the image array nearest the ground plane by the lattice constant. For this simulations the distance between the source and the array of scatterers is $H_y=1$ m, and the site of the receiver is (2.5, 0) m. The position of the source is $(0, 0)$.

Figure \ref{fig:formulations_comparison1} shows the analytical and numerical predicted IL spectra for these two particular cases, acoustically-hard and completely absorbent ground. From the analytical point of view the effect of the ground has been calculated using IMST or the equivalent doubled structure in free field using MST. From the numerical point of view the effect of the rigid ground has been calculated using the finite element method (FEM). The application of FEM to unbounded domains, as for example the case of the scattering problems, involves a domain
decomposition by introducing an artificial boundary around the obstacle. At the artificial boundary, the discretization can be coupled in various ways to some discrete representation of the analytical solution. In this work, we use the Perfectly Matched Layers (PML) \cite{Berenguer94} to numerically approximate the Sommerfeld conditions (see Eq. \ref{form_radiation}). The commercial software COMSOL Multiphysics 3.5 is used in for the simulations.

\begin{figure}[h]
		\center
		\includegraphics[scale=.4]{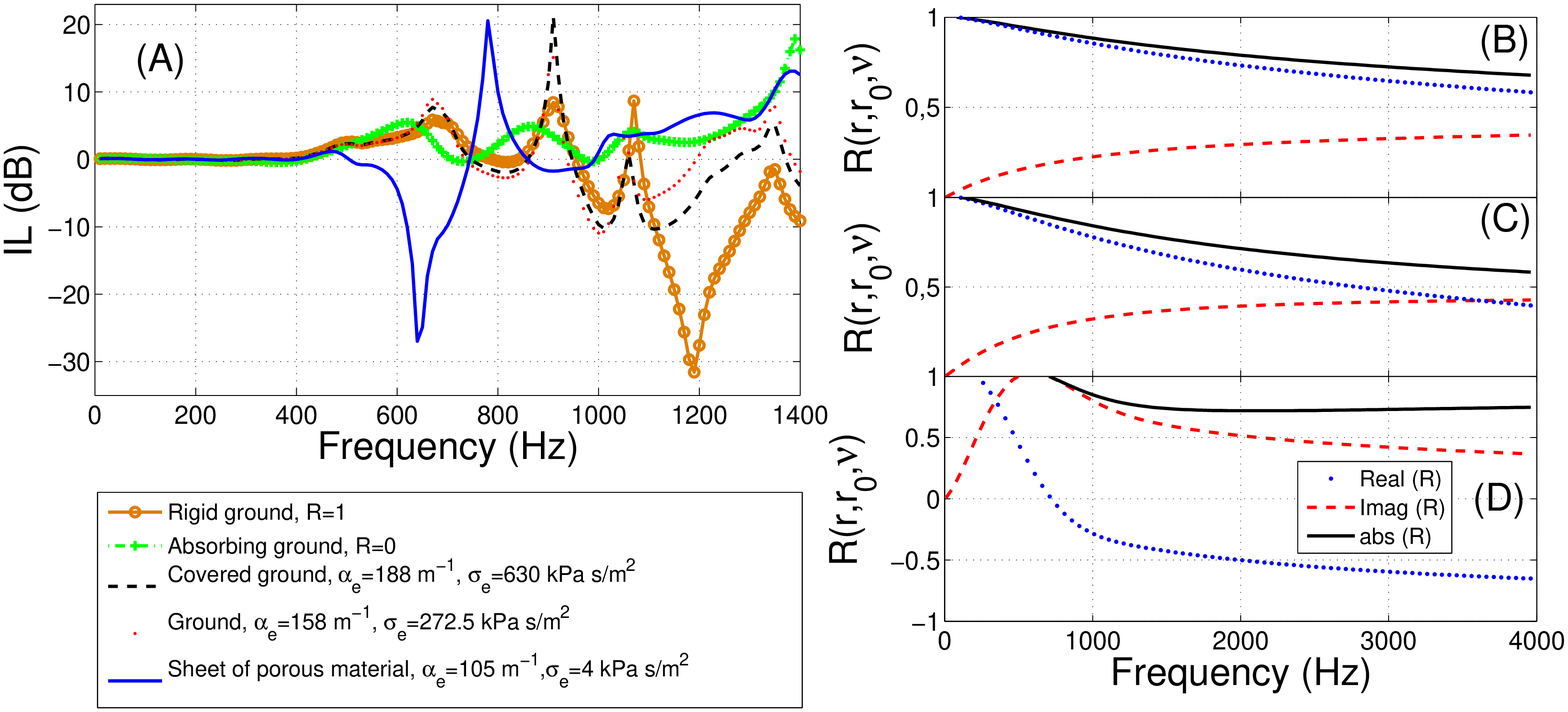}
		\caption{(Color online) IL and reflection coefficient predictions obtained using the IMST for five values of ground impedance. (A) IL of the same array as in Figure \ref{fig:formulations_comparison1}A with the source and the receiver placed at points (0,0.25) m and (2.5, 0.75) m respectively, and $H_x=0.15$ m and $H_y=1$ m. Orange continuous dotted line show the IL predictions for the case of rigid ground. Green crossed dash line shows the IL predictions for the case of completely absorbent ground. Black dashed line shows the IL predictions for the case of a covered ground characterized by $\sigma_e=630$ $\rm kPa\;s/m^2$ and $\alpha_e=188$ $\rm m^{-1}$. Red dots show the IL predictions for the case of a ground characterized by $\sigma_e=272500$ $\rm Pa\;s/m^2$ and $\alpha_e=158$ $\rm m^{-1}$. Blue continuous line shows the IL predictions for the case of a sheet of porous material characterized by $\sigma_e=4000$ $\rm Pa\;s/m^2$ and $\alpha_e=105$ $\rm m^{-1}$. The real, imaginary and absolute values of the reflection coefficients of these three grounds are shown in (B), (C) and (D) respectively.}
		\label{fig:formulations_comparison3}
\end{figure}

In Figure \ref{fig:formulations_comparison1}A the case of the rigid ground is analyzed. Blue continuous line shows the IL produced by a $5\times3$ square array of rigid cylinders with diameter 0.1 m above an acoustically-hard plane at $y = 0$ calculated using IMST. Red open circles show the IL produced by a doubled (i.e. a $10\times3$) lattice in the free field calculated using MST. Moreover, these two cases have been also numerically solved by FEM. Green dots show the IL produced by a $5\times3$ square array of rigid cylinders with diameter 0.1 m above an acoustically-hard plane at y = 0, in which Neumann conditions have been considered. Black pentagrams show the IL produced by a doubled (i.e. a $10\times3$) lattice in the free field. Note the complete agreement between both cases, the red open circles completely coincides with the blue line, and the green dots completely coincide with the black pentagram as it was predicted few lines above. The analytical and numerical predictions are in good agreement taking into account that the possible small differences at high frequencies are produced by the precision of the mesh in the numerical discretization. The differences in the predictions at higher frequencies can be reduced by taking more elements on the solution domain. This however increases the computational time. Because of the low filling fraction of the array (8\%), the pseudogap at $\Gamma X$  direction (0$^\circ$ of incidence) presents low values of insertion loss, however the range of frequencies (490, 609) Hz is in complete agreement with that obtained using plane wave expansion in the band structure shown in Figure \ref{fig:formulations_comparison1}C. The peaks of IL at high frequencies are not in agreement with the band structure. It can be done to the low filling fraction and the low number of rows in the array.

In Figure \ref{fig:formulations_comparison1}B the case of the completely absorbent ground (R=0) with the source on it is analyzed. Blue continuous line shows the IL produced by a $5\times3$ square array above the absorbing ground using IMST. Red open circles show the IL produced by an $5\times3$ array in the free field calculated using MST. Note the complete agreement between both cases.

When the source is not located on the ground, the problem only can be analytically solved using the IMST. We have analyzed the same array as before but varying the impedance of the ground. The finite impedances are characterized by a two parameter impedance model \cite{Taherzadeh} with different values of $\sigma_e$ and $\alpha_e$. We have specifically analyzed the cases of a rigid ground, a completely absorbent ground, a covered ground characterized in reference \cite{Attenborough92}, the finite impedance ground analyzed in reference \cite{Taherzadeh} and the sheet of porous material used in this work with $\sigma_e=4000$ $\rm Pa\;s/m^2$ and $\alpha_e=105$ $\rm m^{-1}$.

Figure~\ref{fig:formulations_comparison3} shows the IL predictions obtained using the IMST for the five values of ground impedance. Figure \ref{fig:formulations_comparison3}A shows the IL obtained but with the source and the receiver placed at points (0,0.25) m and (2.5, 0.75) m respectively. Figures \ref{fig:formulations_comparison3}B, \ref{fig:formulations_comparison3}C and \ref{fig:formulations_comparison3}D show the absolute, real and imaginary values of the reflection coefficients calculated at the receiver site for the three considered grounds impedance based on a two parameter model: $\sigma_e=630$ $\rm kPa\;s/m^2$ and $\alpha_e=188$ $\rm m^{-1}$, $\sigma_e=272500$ $\rm Pa\;s/m^2$ and $\alpha_e=158$ $\rm m^{-1}$, $\sigma_e=4000$ $\rm Pa\;s/m^2$ and $\alpha_e=105$ $\rm m^{-1}$ respectively. 

Orange line in Figure \ref{fig:formulations_comparison3}A shows the IL of the array over a rigid ground. Around 1200 Hz the values of the IL are negative because of the excess attenuation. This effect can be changed as the impedance is considered in the model as it will be seen in Section \ref{sec:results}. Black dashed, red dots, green dot dashed and blue continuous lines show the reduction or shift of the excess attenuation peaks due to the reduction on the reflection coefficient because of the finite impedance on the ground.

\begin{figure}[h]
		\center
    \caption{(Color online) Pressure maps for an array of $7\times 3$ scatterers in square array with lattice constant $a=0.069$ m considering an acoustically-rigid ground with a line source placed at point $O=(0, 0.235)$ m. The circular scatterers present a radius $r=0.0275$ m. For this simulations we consider $H_y=0.755$ m  and $H_x=0.0275$ m. (A) Band structure calculated using PWE. (B) and (C) pressure maps ($Re(P)$) (Pa) at 2000 Hz for an acoustically-hard ground and an finite impedance ground ($\sigma_e=4000$ $\rm Pa\;s/m^2$ and $\alpha_e=105$ $\rm m^{-1}$) respectively.}
    \label{fig:formulations_comparison4}
\end{figure}

In Figure \ref{fig:formulations_comparison4} we have analyzed the symmetry of the acoustic pressure field respect to the ground plane depending on the value of the impedance. If the impedance is infinite, i.e., acoustically-hard ground, the acoustic field should be symmetric, which means that the acoustic field in the image space should be symmetrically equivalent to the acoustic field in the real space. However, in the case of a ground with finite impedance, the symmetry is broken and the field in the real space is not symmetrically equivalent to the acoustic field in the image space. Let us study these differences in the acoustic field considering an array of $7\times 3$ scatterers in square periodicity with lattice constant $a=0.069$ m over a ground with a line source placed at point $O=(0, 0.235)$ m. The circular scatterers present a radius $r=0.0275$ m. For these simulations we consider $H_y=0.755$ m  and $H_x=0.0275$ m. We have used here the configuration defined in the experimental setup presented in Section \ref{experiment}.

Figure \ref{fig:formulations_comparison4}A shows the band structure for this array. Note that for this case, the filling fraction is bigger and the lattice constant is lower than in the previous array. These properties produce both a full band gap and an increasing of the frequencies of the band gap respectively. The range of frequencies between 2478 Hz and 3171 Hz defines the full band gap for this array. Using the analytical method shown in Section \ref{IMST}, IMST, we have predicted the acoustic field 2000 Hz considering both an acoustically-hard ground and a finite impedance ground characterized by a two parameter model ($\sigma_e=4000$ $\rm Pa\;s/m^2$ and $\alpha_e=105$ $\rm m^{-1}$). Figures \ref{fig:formulations_comparison4}B and \ref{fig:formulations_comparison4}C show the pressure maps ($Re(P)$) at 2000 Hz considering an acoustically-hard ground and the finite impedance ground respectively. One can observe clearly the symmetry of the acoustic pressure field for the hard ground case, and the non symmetry in case of finite impedance.

\section{Results}
\label{sec:results}
\subsection{Bandgap-ground plane interaction}

In this Section we analyze the effect of the ground on the attenuation properties of a periodic array of scatterers. For a predetermined source position and impedance of the ground we analyze the Insertion loss defined in this case as the difference between the pressure level measured with ground and the pressure level measured with ground and array. The excess attenuation depends on the position of the receiver and on the frequency. On the other hand, the bandgap of the array of scatterers depends on the position of the source and the receiver and on the filling fraction. Here we consider two different grounds: an acoustically-rigid ground and a finite impedance ground characterized by a two parameter model ($\sigma_e=4000$ $\rm Pa\;s/m^2$ and $\alpha_e=105$ $\rm m^{-1}$). The array considered in this Section is a $7\times 3$ lattice with square periodicity with $a=0.069$ m with a line source placed at point $O=(0, 0.235)$ m. The circular scatterers present a radius $r=0.0275$. For these simulations we consider $H_y=0.755$ m  and $H_x=0.0275$ m. We calculate the IL spectra using IMST for the heights in the interval $y=[0, 0.469]$ m for a distance from the source $x=1.203$ m.

\subsubsection{Hard ground}
The excess attenuation for several heights of the receiver due to a rigid ground can be observed in Figure \ref{fig:results_1}A. Each horizontal cut, $y=y_r$, of the map in Figure \ref{fig:results_1}represents the spectrum at point $(1.203,y_r)$ m. The pressure level in the receiver sites is characterized with the following expression,
\begin{equation}
\label{eq:EA}
PL=20\log_{10}\left(H_0(kr)+H_0(kr')\right).
\end{equation}
The excess attenuation appears in Figure \ref{fig:results_1}A as regions of frequencies with negative values of the PL produced by the destructive interference between the incident wave (from the source) and the reflected wave (reflected in the ground). Then positive values of pressure level means a positive interference and consequently a reinforcement. Negative values means excess attenuation. Figure \ref{fig:results_1}A shows the dependence of the excess attenuation on the height of the receiver and on the frequency. The higher the height, the lower the frequency of the excess attenuation peak. Excess attenuation peaks of second order can be also observed for high values of both heigh and frequencies.
  
\begin{figure}[h]
		\center
    \caption{(Color online) (A) Pressure level spectra surface produced by the line source in presence of a rigid ground. The line source is placed at point $O=(0, 0.235)$ m. We calculate the pressure level using Equation \ref{eq:EA} for the heights in the interval $y=[0, 0.469]$ m for a distance from the source $x=1.203$ m. (B) IL map produced by a $7\times 3$ array with square periodicity with $a=0.069$ m. The circular scatterers present a radius $r=0.0275$. For these simulations we consider $H_y=0.755$ m  and $H_x=0.0275$ m. We calculate the IL spectra using IMST for the heights in the interval $y=[0, 0.469]$ m for a distance from the source $x=1.203$ m. Vertical dashed line marks the beginning of the pseudogap at $\Gamma X$ direction (0$^\circ$) and the vertical continuous line marks the ranges of frequencies of the full bandgap. Horizontal dotted lines show the analytical and experimental cuts shown in Figure \ref{fig:data_vs_MST_hard}.}
    \label{fig:results_1}
\end{figure}

In Figure \ref{fig:results_1}B the IL map produced by the interaction of the rigid ground and the array of scatterers is shown.  The vertical dotted line shows the beginning of the pseudogap at $\Gamma X$ direction whereas the vertical continuous lines show the range of frequencies of the full band gap of the array. This IL is calculated using Equation \ref{form_IL}. In order to understand the meaning of the IL calculated with this formula, one should compare the results of the pressure level in the receiver site without array (Figure \ref{fig:results_1}A) with the results of the IL. 

\subsubsection{Soft ground}
\begin{figure}[h]
		\center
    \caption{(Color online) (A) Pressure level spectra surface produced by the line source in presence of a finite impedance ground ($\sigma_e=4000$ $Pa\;s/m^2$ and $\alpha_e=105$ $m^{-1}$). The line source is placed at point $O=(0, 0.235)$ m. We calculate the pressure level using Equation \ref{eq:EA_2} for the heights in the interval $y=[0, 0.469]$ m for a distance from the source $x=1.203$ m. (B) IL map produced by a $7\times 3$ array with square periodicity with $a=0.069$ m above a finite impedance ground. The circular scatterers present a radius $r=0.0275$. For these simulations we consider $H_y=0.755$ m  and $H_x=0.0275$ m. We calculate the IL spectra using IMST for the heights in the interval $y=[0, 0.469]$ m for a distance from the source $x=1.203$ m. Vertical dashed line marks the beginning of the pseudogap at $\Gamma X$ direction (0$^\circ$) and the vertical continuous line marks the ranges of frequencies of the full bandgap. Horizontal dotted lines show the analytical and experimental cuts shown in Figure \ref{fig:data_vs_MST_soft}.}
    \label{fig:results_2}
\end{figure}

The excess attenuation for several heights of the receiver due to a soft ground without array of scatterers can be observed in Figure \ref{fig:results_2}A. The pressure level in the receiver sites is characterized with the following expression,
\begin{equation}
\label{eq:EA_2}
PL=20\log_{10}\left(H_0(kr)+R(\vec{r}_O,\vec{r}_R;\nu)H_0(kr')\right),
\end{equation}
where $R(\vec{r}_O,\vec{r}_R;\nu)$ is calculated using the approach shown in Section \ref{sec:ground}. The impedance of the ground considered in this work is characterized by a two parameter model ($\sigma_e=4000$ $\rm Pa\;s/m^2$ and $\alpha_e=105$ $\rm m^{-1}$) and the reflection coefficient can be obtained combining Equation \ref{eq:R} in Equation \ref{impedance}. 

In Figure \ref{fig:results_2}A one can observe the pressure level for the case of a sound source with this impedance ground. Again the excess attenuation also depends on the frequency and on the heigh of the receiver. However, the dependence on this parameters is changed because of the properties of the ground. The first excess attenuation peak appears at lower frequencies and lower heights than in the case of acoustically-rigid ground. Excess attenuation peaks of second order can be also observed for high values of both heigh and frequencies with lower values than in the case of rigid ground. 

In Figure \ref{fig:results_2}B the IL maps produced by the interaction of the soft ground and the array of scatterers is shown.  The vertical dotted line shows the beginning of the pseudogap at $\Gamma X$ direction whereas the continuous lines show the range of frequencies of the full band gap of the array. In the case of the IL of the array over the soft ground the attenuation peaks due to the interaction between the ground are also present. One can also observe that the array of scatterers can change the attenuation properties at the receiver site for frequencies above and below the first peak of excess attenuation adding the effect to that related with the bandgap.

\subsection{Comparisons between data and predictions}\label{comparisons}

\subsubsection{Laboratory experiment}\label{experiment}

2 m long PVC cylinders with outer diameters of 55 mm have been used to construct the periodic array with lattice constant$a=0.069$ m. The sound source was a Bruel $\&$ Kjaer point source loudspeaker controlled by a Maximum-Length Sequence System Analyzer (MLSSA) system enabling determination of impulse responses. Measurements of the insertion loss (IL) spectra for arrays of cylinders placed near to a ground surface in an anechoic chamber have been obtained. 


A 0.03 m thick wooden board large enough to avoid the diffraction at the edges was used as a hard surface. The loudspeaker point source was positioned 0.755 m from the array at the height of the horizontal mid-plane of the array (0.23 m above the ground). The height of the receiver microphone was 0.117 m, 0.235 m or 0.352 m and it was placed in a vertical plane 0.257 m from the back of the array. The receiver heights were chosen to be below, at, and above, the horizontal mid-plane of the array. In all cases, a constant distance between the microphone and the periodic array has been considered, in such a way that the distance between the source and the receiver is $x=1.203$ m. The difference between the sound levels recorded in the  X direction ($0^{\circ}$) at the same point with and without the ground plus the array was measured. 

\begin{figure}[h]
		\center
		\caption{Measured (open circle with dotted line) and predicted (continuous line) insertion loss spectra for source at coordinates (0,0.235) m and 0.755 m from a $7\times3$ array of rigid cylinders of diameter 0.055 m over acoustically-hard ground with receiver coordinates (A) (1.203,0.117) m, (B) (1.203,0.235) m and (C) (1.203,0.352) m. Arrays of cylinders placed near to a ground surface.}
		\label{fig:data_vs_MST_hard}
\end{figure}

\subsubsection{Hard ground}
Figure~\ref{fig:data_vs_MST_hard} compares measured and predicted IL spectra for $7\times3$ rigid cylinder arrays above a hard ground plane for three receiver heights using the source location described in Section \ref{experiment}. The agreement between predictions and measurements is fairly good. The analytical spectra corresponding to these three heights are also marked in Figure \ref{fig:results_1}B with dotted horizontal lines. One can also observe the ground effect on the IL due to the excess attenuation peaks for the three heights analyzed in this work near  4000 Hz, 2000 Hz and 1400 Hz respectively. The horizontal dotted lines in Figure \ref{fig:results_1}B shows the corresponding cuts of the IL maps for the three heights analyzed in this Section.

\subsubsection{Soft ground}
Figure \ref{fig:data_vs_MST_soft} compares corresponding measured and predicted insertion loss spectra for $7\times3$ rigid cylinder arrays over finite impedance ground for three receiver heights using the source location described in section~\ref{experiment}. Again the agreement between predictions and measurements is fairly good. The adverse and the additional attenuation influences of ground effect on the IL spectra are shifted towards lower frequencies because of the finite impedance of the ground. The horizontal dotted lines in Figure \ref{fig:results_1}B shows the corresponding cuts of the IL maps for the three heights analyzed in this Section.

\begin{figure}[h]
		\center
		\caption{Measured (solid line) and predicted (broken line) insertion loss spectra for source at coordinates (0,0.235) m and 0.755 m from a $7\times3$ array of rigid cylinders of diameter 0.055 m over finite impedance ground with receiver coordinates (a) (1.203,0.117) m, (b) (1.203,0.235) m and (c) (1.203,0.352) m. Arrays of cylinders placed near to a ground surface.}
		\label{fig:data_vs_MST_soft}
\end{figure}

\section{Concluding remarks}\label{conclusion}
The effect of both rigid and finite impedance grounds on the attenuation properties of an array of rigid cylindrical scatterers has been analytically and experimentally analyzed. The Image Multiple Scattering Theory (IMST) have been developed as an analytical methodology to study the effect of several finite impedance grounds on the propagating properties of an array of rigid scatterers in air. The dependence of the attenuation properties for several heights on the attenuation has been analyzed in this paper by means of IMST and experimentation. The excess attenuation produced by the interaction of the sound source and the ground can be used to reduce or increase the attenuation properties of the array of scatterers. On one hand the deeps on the excess attenuation always increases the attenuation properties of the array of rigid scatterers. On the other hand, the regions of frequencies out of the excess attenuation, produces adverse influences in the attenuation of the array at the receiver positions. Then the excesses attenuation should be taken into account in the design of array of scatterers acting as sonic crystal noise barrier. 

\section*{Acknowledgment}
Authors would like to acknowledge the Open University by the facilities for the measurements. This work was supported by MEC (Spanish Government) and FEDER funds, under Grant No. MAT2009-09438.


\begin{thebibliography}{99}
\bibitem{Sanchez-PerezA}
J.V. S\'anchez-P\'erez, C. Rubio, R. Mart\'inez-Sala, R. S\'anchez-Grandia and V. G\'omez, ``Acoustic barriers based on periodic arrays of scatterers'', Appl. Phys. Lett., \textbf{81}, 5240 (2002).

\bibitem{Sanchez10}
J.~S\'anchez-Dehesa, Victor~M. Garcia-Chocano, D.~Torrent, F.~Cervera, and
  S.~Cabrera. Noise control by sonic crystal barriers made of recycled materials.
arXiv:1004.2570v1 [cond-mat.mtrl-sci], (2010).

\bibitem{SUOU}
A. Krynkin, O. Umnova, A.Y.B. Chong, S. Taherzadeh, and K. Attenborough, ``Predictions and measurements of sound transmission through a periodic array of elastic shells in air'', J. Acous. Soc. Am., accepted.

\bibitem{Sanchez-PerezB}
J. V. S\'anchez-P\'erez, D. Caballero, R. Mart\'inez-Sala, C. Rubio, J. S\'anchez-Dehesa, F. Meseguer, J. Llinares, and F. G\'alvez, Phys. Rev. Lett. \textbf{80}, 5325 (1998).

\bibitem{Liu00a}
Z. Liu and X. Zhang and Y. Mao and Y.Y. Zhu and Z. Yang and C.T.
	Chan and P.Sheng
``Locally Resonant Sonic Materials'' Science, \textbf{289}, 1734, (2000).

\bibitem{Umnova}
O. Umnova, K. Attenborough and C.M. Linton, ``Effects of porous covering on sound attenuation by periodic arrays of cylinders'',  J. Acous. Soc. Am., \textbf{119}, 278--284 (2006).

\bibitem{Romero06}
V.~Romero-Garc\'ia, E.~Fuster, L.~M. Garcia-Raffi, E.~A. S\'anchez-P\'erez,
  M.~Sopena, J.~Llinares, and J.~V. S\'anchez-P\'erez.
``Band gap creation using quasiordered structures based on sonic crystals''. Appl. Phys. Lett., \textbf{88},174104, (2006).


\bibitem{Castineira10}
S.~Casti{\~n}eira-Ib{\'a}{\~n}ez, V.~Romero-Garc{\'i}a, J.~V.
  S{\'a}nchez-P{\'e}rez, and L.~M. Garcia-Raffi. ``Overlapping of acoustic bandgaps using fractal geometries'', Eur. Phys. Let., \textbf{92}, 24007, (2010).

\bibitem{Zaviska}
F. Zaviska, ``The deflection of electro magnetic waves on parallel, infinite long orbital cylinder'', Annalen der Physik, \textbf{40}, 1023--1056 (1913).

\bibitem{Linton}
C.M. Linton and D.V. Evans, ``The interaction of waves with arrays of vertical circular cylinders'', Journal of Fluid Mechanics, \textbf{215}, 549--569 (1990).

\bibitem{Martin}
P.A. Martin, Multiple Scattering: Interaction of Time-Harmonic Waves with N Obstacles, Cambridge University Press, Cambridge, 2006.

\bibitem{Chen01}
You~Yu Chen and Zhen Ye.``Theoretical analysis of acoustic stop bands in two-dimensional periodic scattering arrays'',
 Phys. Rev. E, \textbf{64}, 036616, (2001).

\bibitem{Allen79}
Jont~B. Allen and David~A. Berkley, ``Image method for efficiently simulating small-room acoustics'',
J. Acoust. Soc. Am., \textbf{65}, 943--950, (1979).

\bibitem{Boulanger}
P. Boulanger, K. Attenborough, Q. Qin and C.M. Linton, ``Reflection of sound from random distributions of semi-cylinders on a hard plane: models and data'', Journal of Physics D: Applied Physics, \textbf{38}, 3480--3490 (2005).

%


%

\bibitem{Taherzadeh}
S. Taherzadeh and K. Attenborough, ``Deduction of ground impedance from measurements of excess attenuation spectra'', J. Acous. Soc. Am., \textbf{105}, 2039--2042 (1999)

\bibitem{Piercy77}
J.~E. Piercy, T.~F.~W. Embleton, and L.~C. Sutherland.
 ``Review of noise propagation in the atmosphere''.
J. Acoust. Soc. Am., \textbf{61}, 1403--1418, (1977).

\bibitem{Attenborough92}
Keith Attenborough, ``Ground parameter information for propagation modeling'', J. Acoust. Soc. Am, \textbf{92}, 418--427, (1992).

\bibitem{Attenborough95}
Keith Attenborough and Shahram Taherzadeh. ``Propagation from a point source over a rough finite impedance boundary'', J. Acoust. Soc. Am, \textbf{98}, 1717--1722, (1995).

\bibitem{Berenguer94}
J.P. Belenguer, ``A perfectly matched layer for the absorption of electromagnetic waves'',
 J. Compt. Physics, \textbf{114}, 185, (1994).

\bibitem{Abramowitz}
M. Abramowitz and I.A. Stegun, Handbook of Mathematical Functions, National Bureau of Standards, Washington, 1964, p. 255. 

%
%

\end{thebibliography}
\end{document}